\documentclass[twoside]{article}
\usepackage{fleqn,espcrc2}
\usepackage{graphicx}
\usepackage[figuresright]{rotating}
\title{C-axis optical properties of high T$_c$ cuprates\thanks{
Submitted to Physica C.Invited talk at the 6th International 
Conference on Materials and Mechanisms of Superconductivity and 
High Temperature Superconductors (M2SHTSC), February 20-25 2000, Houston, USA.}}
\author{D. van der Marel, A. Tsvetkov, M. Grueninger, D. Dulic, H. J. A. Molegraaf
\address{Laboratory of Solid State Physics\\ Materials Science Centre,
Nijenborgh 4, 9747 AG Groningen, The Netherlands}
\thanks{This investigation was supported by the Netherlands Foundation for 
Fundamental Research on Matter (FOM) with financial aid
from the Nederlandse Organisatie voor Wetenschappelijk Onderzoek (NWO).}}       
\begin{document}
\begin{abstract}
A review is given of the experimental status of the interlayer
coupling energy in the cuprates. A second c-axis plasmon is
identified in the double layer compound Y123 for various dopings.
The anomalous transport properties along the c-direction and in the
planar directions are compared to model calculations based on
strongly anisotropic scattering. An excellent description of 
the optical data at optimal doping is obtained if an anomalously
large anisotropy of the scattering rate between cold spots
and hot spots is assumed. This raises questions as to the 
physical meaning of these parameters.
\vspace{1pc}
\end{abstract}
\maketitle
\begin{figure}[htb]
\vspace{9pt}
\includegraphics*[width=70mm]{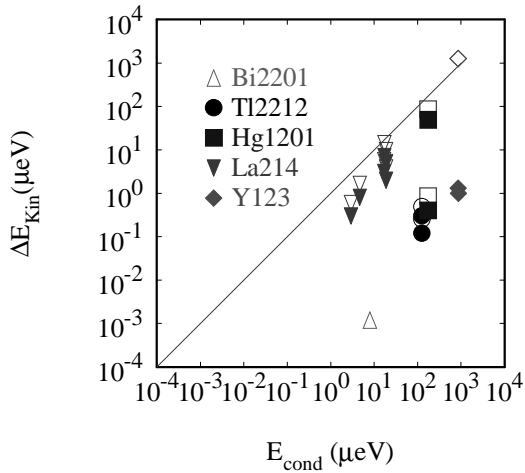}
  \caption{C-axis kinetic 
  energy\cite{kim94,kam98,ilt98,basov99,kirtley99,matsuda99,shibauchi94,pana97,gru2000} 
  versus condensation energy\cite{loram}.The open symbols represent the most
  E$_J$ estimated from either the JPR or the c-axis penetration depth. The closed 
  symbols represent the difference in low energy spectral weight between the 
  superconducting and the normal state.}
  \label{econd}
\end{figure}
\begin{figure}[htb]
\vspace{9pt}
\includegraphics*[width=70mm]{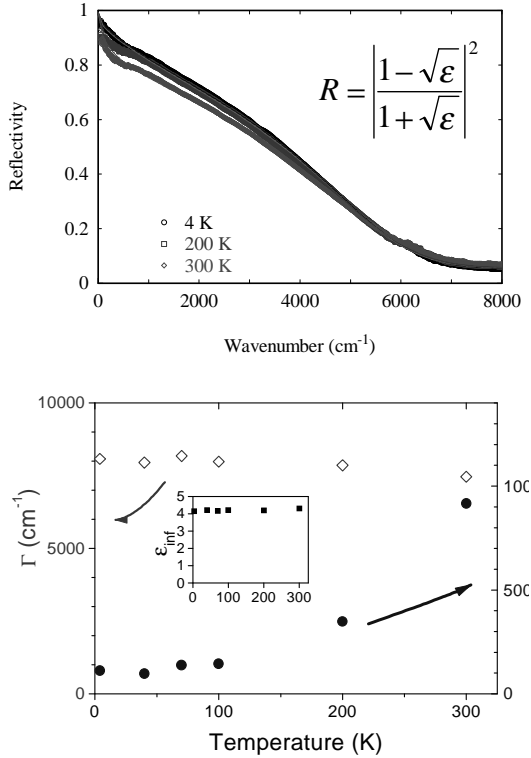}
  \caption{Top: Reflectivity curves of Bi2201 adopted from Ref.\cite{bi2201} 
  (open symbols), and fits to the anisotropic scattering model (solid curves). 
  Bottom: Fitparameters, $\omega_p/2\pi c=13700$cm$^{-1}$.} 
  \label{reflect}
\end{figure}

During the past four years the issue whether the mechanism of superconductivity in the
cuprates could be a lowering of {\em kinetic} energy (as opposed to {\em potential}
energy in BCS theory) has received considerable attention both theoretically
\cite{hirsch92,pwa95,ajl96,chakra99} and 
experimentally\cite{ilt96,pana97,ilt98,kam98,basov99,kirtley99,matsuda99}. 
Although originally conceived as an {\em in-plane} 
mechanism in the hole-model of superconductivity\cite{hirsch92}, attention later
was concentrated on the c-axis properties\cite{pwa95} first of all because the 
c-axis transport of quasiparticles had been found to have a very large
scattering rate in the normal state\cite{lance} and,                      
rather surprisingly, {\em also} in the superconducting state\cite{kim94}, thus 
providing a channel for kinetic energy lowering for paired charge carriers as soon as they
become delocalized as a result of the pairing. A high value of the scattering rate
for transport along the c-direction which remains high in the superconducting state
appears to be a robust property of the cuprates: It has been reported for  
La214\cite{kim94}, Y123\cite{hosseini} Tl2201, and Tl2212 \cite{dulic99}.
In the second place the kinetic
energy lowering is just the Josephson coupling energy (or in any case not larger)
in the interlayer tunneling (ILT) model, which suggested a direct experimental way to test the model by
measuring both the condensation energy ($E_{cond}$) and E$_J$. The ILT hypothesis
requires that $E_J\approx E_{cond}$. To avoid the complexity of having {\em two}
possible Josephson junctions per unit cell of different strength, single layer
cuprates had to be considered. Among those Tl2201 had one of the highest T$_c$'s
($\simeq 80$ K), and relatively large (though thin along the $c$-direction) crystals
and thin films were available. In the spring of 1996 the first experimental
results were presented\cite{ilt96},
showing that $E_J$ was at least two orders of magnitude too small to account for
the condensation energy. Although these results seemed to rule out ILT as the main
mechanism of superconductivity\cite{ajl96}
they relied on the non-observance of a plasma-resonance where it should have been in
the superconducting state (800 cm$^{-1}$). The issue remained dormant until 
first $\lambda_c$\cite{kam98} of 17 $\mu$m and next the Josephson plasma resonance
(JPR)\cite{ilt98} at $28$ cm$^{-1}$ had
been observed experimentally, allowing a precise determination of $E_J\approx 0.3 \mu$eV
in Tl2201 with $T_c=80$k. This is a factor 400 lower than $E_{cond}\approx 100 \mu$eV per
copper, based either on $c_V$ experimental data\cite{loram}, or on the formula
$E_{cond}=0.5N(0)\Delta^2$ with $N(0)=1 eV^{-1}$ per copper, and $\Delta\simeq 15 meV$.
A c-axis kinetic energy change even {\em smaller} than $E_J$ is obtained from estimating
the amount of high energy spectral weight transferred to the $\delta$-function at
zero frequency\cite{basov99}: In the examples studied so far this gives
a value of $\Delta E_{kin,c}$ wich is 0.5 $E_J$, for the underdoped materials,
and less than 0.1 $E_J$ for the optimally doped materials. 
In Fig. \ref{econd} the change in c-axis kinetic energy and the Josephson
coupling energies are compared to the condensation energy for a large number
of high T$_c$ cuprates. For most materials materials we see, that
$E_J < E_{cond}$, sometimes differing by several orders of magnitude. 
\\
In this plot we have also indicated optimally doped and overdoped Y123. 
Below T$_c$ we observe a transfer of spectral
weight from the FIR not only to the condensate at $\omega $=0, but also to a new peak
in the MIR. This peak is naturally explained\cite{gru2000,munzar} 
as a transverse out-of-phase bilayer plasmon\cite{dbleplas} by a model for $\sigma(\omega)$ which 
takes the layered crystal structure into account. With decreasing doping the 
plasmon shifts to lower frequencies and can be identified with the surprising 
and so far not understood FIR feature reported in {\it underdoped} bilayer cuprates.
A second Josephson plasmon\cite{dbleplas} has also been reported\cite{shibata} 
for the T$^*$ phase La$_{1-x}$Sr$_x$SmCuO4. For points marked YBCO $\Delta E_{kin}$
was calculated from the total superfluid spectral weight of the two plasmons.
For optimally doped and overdoped YBCO almost all (at least 95 $\%$)
superfluid spectral weight originates from the gap-region, resulting
in the solid points.
\\
As mentioned in the beginning of this paper, there is the issue of 
the very large scattering rate in the normal state\cite{lance} and,                      
rather surprisingly, {\em also} in the superconducting 
state\cite{kim94,hosseini,dulic99}. Usually a large scattering rate
along the c-axis is interpreted as a form of tunneling with a large
scattering of $k_{\parallel}$ of the charge carriers. The
term 'incoherent' is usually reserved for non- $k_{\parallel}$
conserving tunneling. Clearly there must be some degree of 
$k_{\parallel}$ conservation in the tunneling, as otherwise the
c-axis critical current would be zero due to cancellation of the
phases of the d-wave order parameter. However, another form of incoherent transport
exists, namely where $k_{\parallel}$ is conserved, while  the
memory of $k_{\perp}$ is lost on the timescale of a tunneling event.
{\em If} c-axis tunneling is $k_{\parallel}$-conserving, this has a number of 
interesting consequences. 
\\
In the first place the tunneling matrix elements depend
strongly on $k_{\parallel}$: As a result of some peculiarities of the
crystal structure of these materials it has zero's in the zone-diagonal
directions\cite{chakravarty}. 
\\
There are indications that the charge carrier scattering rate is also strongly 
$k_{\parallel}$  dependend, probably due to coupling to 
spin-fluctuations\cite{pines}: The zone-diagonal directions remain
unaffacted, while the $(\pi,0)$ directions have a strong scattering. 
\\
This leads to a simple formula for the in-plane optical conductivity
in the normal state\cite{millis,marel99}
$\epsilon_{ab}(\omega)=\epsilon_{\infty}-\frac{\omega_p^2\tau}{\omega\sqrt{1-i\omega\tau}
\sqrt{1+\Gamma\tau-i\omega\tau}}$.
Here $\Gamma$ is the hot-spot scattering rate, $1/\tau$ is the cold-spot
scattering rate, and the above expression was derived assuming that the
scattering rate varies smoothly between these two extrema along the 
Fermi-surface. In Fig. \ref{reflect} we provide reflectivity curves of Bi2201 
(T$_c \approx 10 K$) taken from Ref. \cite{bi2201} together with the
four parameter fits. In the fit procedure the value of $\omega_p$ was
kept fixed at 13700 cm$^{-1}$ at all temperatures, while $\epsilon_{\infty}$,
$\tau$ and $\Gamma$ were adjusted to obtain the best fit. It turned out, 
that $\epsilon_{\infty}=4.2\pm0.1$ at all temperatures. The temperature dependence
of $\Gamma$ and $1/\tau$ are indicated in the lower panel of Fig. \ref{reflect}. We see,
the model leads to a {\em very} large anisotropy between these two scattering
rates: $\Gamma$ is almost a constant, while $1/\tau$ has a $T^2$ temperature
dependence on top of a small residual value. 
%
%
\begin{figure}[htb]
\vspace{9pt}
\includegraphics*[width=70mm]{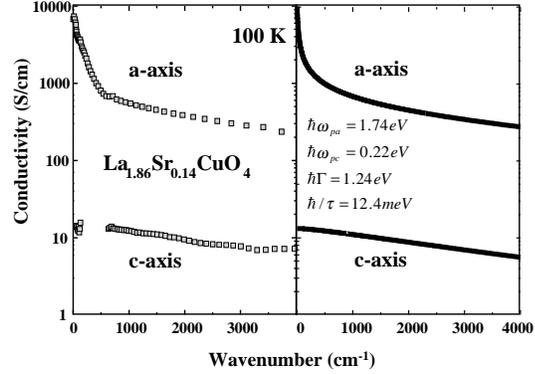}
  \caption{Left:Experimental $\sigma_{ab}$ and $\sigma_c$ of 
  La214 (T$_c$ = 32 K),
  adopted from Refs\cite{somal96} and \cite{gruphd}.
  Right: Comparison of $\sigma_{\parallel}$ and $\sigma_c$ using the model
  expressions of \cite{marel99}, and using the parameters
  $\hbar\omega_{p,a}=1.7 eV$,
  $\hbar\omega_{p,c}=0.2 eV$, $\hbar\Gamma=1.2 eV$, and $\hbar/\tau=12 meV$.} 
  \label{compac}
\end{figure}
%
%
In fact the parameters obtained with this fit look quite unreasonable. A scattering
rate of almost 1 eV around the hot spots is an order of magnitude larger than
typical linewidths observed with ARPES. On the other hand for the optical
spectra a rather complete and selfconsistent description is obtained: The
optical conductivity along the c-axis is largely determined by the
hot-spots, as a result of the strong k-dependence of $k_{\parallel}$. The
resulting analytical expressions for the c-axis conductivity provide
spectra which closely resemble the experimentally observed optical
conductivity along $c$. In the righthand panel of Fig. \ref{compac} we display the 
theoretical curves for the in-plane and c-axis conductivity using the
same parameters as above. In the lefthand panel of Fig. \ref{compac} the experimental
curves for La214 along the two crystallographical directions
are displayed\cite{somal96,gruphd}.
Clearly there is a close resemblence between these datat sets. The significance
of these results is really not clear at this moment. Questions that need
to be answered are: 
\begin{enumerate}
\item
To what extent is $k_{\parallel}$ conserved in the
tunneling. and possible implications for the theory of transport in the cuprates?
\item
What is the minimum value of $t_{\perp}(k_{\parallel})$? The 'chemical' arguments mentioned
abouve provide no arguments why it should be exactly zero?
\item
Do the minimum value of the hopping parameter, of the scattering
rate and of the gap always coincide at exactly the same value of $k_{\parallel}$?
This is not dictated by the symmetry of the materials, which is more often
than not orthorhombic rather than tetragonal.
\item
If the answer to the above is affirmative, what is the microscopic reason?
\item
Why are the scattering rate observed with ARPES and transport/optical probes
completely different?
\end{enumerate}

\end{document}